\documentclass[twocolumn,showpacs,amsmath,amssymb,prd,epsf]{revtex4}

\usepackage{graphicx}
\usepackage{graphicx}% Include figure files
\usepackage{dcolumn}% Align table columns on decimal point
\usepackage{bm}% bold math
\usepackage{epsf}

\def\agt{\mathrel{\raise.3ex\hbox{$>$}\mkern-14mu\lower0.6ex\hbox{$\sim$}}}
\def\alt{\mathrel{\raise.3ex\hbox{$<$}\mkern-14mu\lower0.6ex\hbox{$\sim$}}}

\newcommand{\beq}{\begin{equation}}
\newcommand{\eeq}{\end{equation}}
\newcommand{\beqn}{\begin{eqnarray}}
\newcommand{\eeqn}{\end{eqnarray}}

\begin{document}

\title{Effects of hyperons in binary neutron star mergers}

\author{Yuichiro Sekiguchi}

\author{Kenta Kiuchi}

\author{Koutarou Kyutoku}

\author{Masaru Shibata}

\affiliation{Yukawa Institute for Theoretical Physics,
Kyoto University, Kyoto 606-8502, Japan
}

\begin{abstract}
Numerical simulations for the merger of binary neutron stars are
performed in full general relativity incorporating both nucleonic and
hyperonic finite-temperature equations of state (EOS) and neutrino
cooling for the first time.  It is found that even for the hyperonic
EOS, a hypermassive neutron star is first formed after the
merger for the typical total mass $\approx 2.7M_{\odot}$, and
subsequently collapses to a black hole (BH).  It is shown that
hyperons play a substantial role in the post-merger dynamics, torus
formation around the BH, and emission of gravitational waves (GWs).
In particular, the existence of hyperons is imprinted in GWs.
Therefore, GW observations will provide a potential opportunity to
explore the composition of the neutron star matter.
\end{abstract}
\pacs{04.25.Dm, 04.30.-w, 04.40.Dg}

\maketitle

{\em Introduction}: The properties of the neutron star (NS) matter, in
particular its equation of state (EOS), are still poorly understood.
In recent years, the influence of non-nucleonic degrees of freedom,
such as hyperons, meson condensations, and quarks, on
properties of the NS matter has been discussed
extensively~\cite{exotic,Lattimer}.  From an energetic point of view,
eventual appearances of these {\it exotic} phases seem to be natural
because of the Pauli principle: the chemical potential for neutrons
will eventually exceed that of exotic particles at some high
densities. However, it is currently uncertain at which density and
temperature these exotic particles appear, because their interactions
are not accurately known.  Generally speaking, presence of such exotic
particles results in softening of EOS. This effect reduces the maximum
mass of NS. The recent mass measurement of PSR J1614-2230
($M_{\rm J1614-2230}=1.97 \pm 0.04M_{\odot}$)~\cite{Demorest10} gave a 
strong impact on the properties of the NS matter, indicating that stiff
EOSs are preferable. However, it does not still constrain the
existence of exotic particles in NS~\cite{exotic2,Lattimer}.

There have been a number of numerical studies exploring effects of
such exotic phases on astrophysical
phenomena~\cite{Nakazato,Sagert,Sumiyoshi,Bauswein,Hotokezaka,BJ11},
aiming at extracting information of the NS matter. In the context of
stellar core collapse, effects of hadron-quark phase transition on
dynamics and neutrino signals were studied in \cite{Nakazato,Sagert}
using the MIT bag model for the quark phase. The hadron-quark
transition is imprinted in the neutrino signals, and in special
conditions, it could generate a second shock wave which triggers a
delayed supernova explosion~\cite{Sagert}. 
Sumiyoshi et al.~\cite{Sumiyoshi} studied the effects of hyperons in 
failed supernovae adopting an EOS developed in~\cite{Ishizuka08}. 
By contrast, effects of the exotic particles on dynamics and gravitational
waves (GWs) from binary neutron stars (BNS) mergers and hypermassive
neutron stars (HMNS) subsequently formed have not been studied in
detail (but see recent studies in ~\cite{Bauswein,Hotokezaka}),
although the coalescence of BNS is one of the most promising sources
for next-generation kilo-meter-size GW detectors~\cite{LIGOVIRGO}.

Among exotic particles, $\Lambda$ hyperons are believed to appear
first in (cold) NS around the rest-mass density of $\rho
\sim$2--3$\rho_{\rm nuc}$~\cite{exotic,Ishizuka08}, where $\rho_{\rm
nuc} \approx 2.8 \times 10^{14}$g/cm$^{3}$ is the nuclear matter
density. 
In this {\it Letter}, we present the first results of
numerical-relativity simulations for the BNS merger performed
incorporating a finite-temperature EOS including contributions of
$\Lambda$ hyperons~\cite{ShenHyp} (Hyp-EOS).
We ignore the contributions of $\Sigma$ hyperons because recent 
experiments suggest that $\Sigma$ hyperons feel a repulsive potential \cite{Noumi02} 
while $\Lambda$ hyperons feel an attractive potential \cite{Ishizuka08}, 
and hence, $\Lambda$ hyperons are likely to appear first and to be dominant. 
In the following, we report the effects of $\Lambda$ hyperons on merger dynamics, 
black hole (BH) formation, neutrino signals, and gravitational waveforms. 
We show that the emergence of $\Lambda$ hyperons is imprinted in GWs from the HMNS.

{\em Setting of numerical simulations}: Numerical simulations in full
general relativity are performed using the same method and formulation
as in~\cite{Sekiguchi1,KSST}: Einstein's equations are solved in the
so-called Baumgarte-Shaprio-Shibata-Nakamura (BSSN)-puncture
formulation~\cite{BSSN} with dynamical gauge conditions for the
lapse function and the shift vector; a fourth-order finite
differencing in space and a fourth-order Runge-Kutta time integration
are used; a conservative shock capturing scheme with third-order
accuracy in space and fourth-order accuracy in time is employed for
solving hydrodynamic equations. In addition to the ordinary
hydrodynamic equations, we solve evolution equations for the neutrino
($Y_{\nu}$), electron ($Y_e$), and total lepton ($Y_l$) fractions per
baryon, taking account of weak interaction processes and neutrino
cooling employing a general relativistic leakage scheme for electron
neutrinos ($\nu_e$), electron antineutrinos ($\bar \nu_e$), and other
types ($\mu/\tau$) of neutrinos ($\nu_x$)~\cite{Sekig,SS11}.

The results with Hyp-EOS are compared with those with a
finite-temperature nucleonic EOS~\cite{Shen}
(Shen-EOS). The maximum mass of zero-temperature spherical NS for
Hyp-EOS is $M_{\rm max, Hyp} \approx 1.8M_{\odot}$. This is smaller
than that for Shen-EOS, $M_{\rm max, Shen} \approx 2.2M_{\odot}$.
Although $M_{\rm max, Hyp}$ is slightly smaller than $M_{\rm
J1614-2230}$, Hyp-EOS deserves employing to explore the impact of
hyperons on the BNS merger, because it can be a viable EOS of the
NS matter not in extremely high densities and hence for studying the
evolution of the HMNS that do not have the extremely high densities
for most of their lifetime. Simulations with finite-temperature EOSs 
with exotic phases which can produce a stable NS with mass larger than $M_{\rm J1614-2230}$ 
should be performed in the future studies.

Numerical simulations are performed in a non-uniform
grid~\cite{Sekiguchi1,KSST}.  The inner domain is composed of a finer
uniform grid and the outer domain of a coarser nonuniform grid. The
grid resolution in the inner zone is chosen so that the major diameter
of each neutron star in the inspiral orbit is covered by 60 and 80
grid points for low- and high-resolution runs, respectively, and we
confirm that the convergence is achieved except for stochastic behaviors due to
convective motions.
Outer boundaries are located in a local wave zone (at $\approx 560$--600~km
along each coordinate axis which is longer than gravitational
wavelengths in the inspiral phase).  During the simulations, we check
the conservation of baryon rest-mass, total gravitational mass, and
total angular momentum, and find that the errors are within
0.2\%, 1\%, and 2\%, respectively, for the high-resolution runs.

This {\it Letter} focuses on the merger of equal-mass BNS.  The
gravitational mass of single NS in isolation is $M_{\rm NS}=1.35$ and
$1.5M_{\odot}$ for Hyp-EOS (referred to as H135 and H15).  We perform
simulations with the initial condition of about 3--4 orbits before the
onset of the merger until the system relaxes to a quasistationary
state. Quasiequilibrium states of BNS are prepared as the initial
conditions, as in~\cite{STU2}.  We compare the simulation results with
those for Shen-EOS with $M_{\rm NS}=1.35$, 1.5, and $1.6M_{\odot}$
(referred to as S135, S15, and S16) obtained in \cite{Sekiguchi1}.  We
do not take account of magnetic fields in the present simulations,
which may play a role in the late-phase evolution after the onset of the
merger~\cite{LR}.

\begin{figure}[t]
 \epsfxsize=3.2in
 \leavevmode
 \epsffile{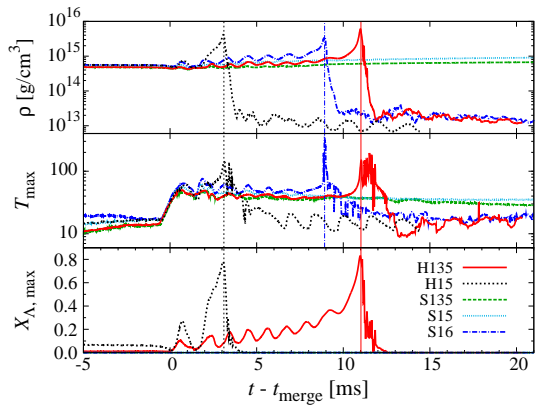}
 \vspace{-3mm}
 \caption{Maximum rest-mass density, maximum matter temperature, and
   maximum mass fraction of hyperons as functions of time for H135 (solid red),
   H15 (dotted black), S135 (dashed green), S15 (short-dotted cyan), and S16 (dashed-dotted blue).
   The vertical thin lines show the time at which a BH is formed.
   \label{fig1}}
\end{figure}

\begin{figure}[t]
 \epsfxsize=3.2in
 \leavevmode
 \epsffile{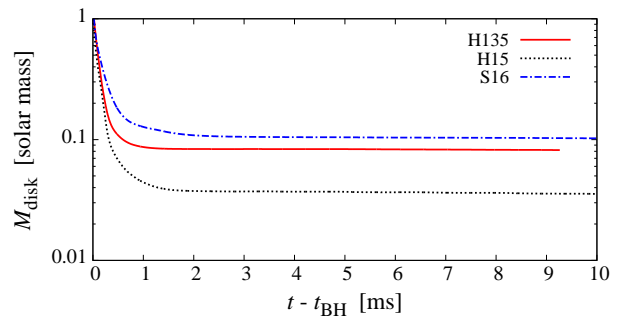}
 \vspace{-3mm}
 \caption{The torus mass as a function of time after the BH formation
   for H135 (solid red), H15 (dotted black), and S16 (dashed-dotted blue).
   \label{fig2}}
\end{figure}

\begin{figure}[t]
 \epsfxsize=3.2in
 \leavevmode
 \epsffile{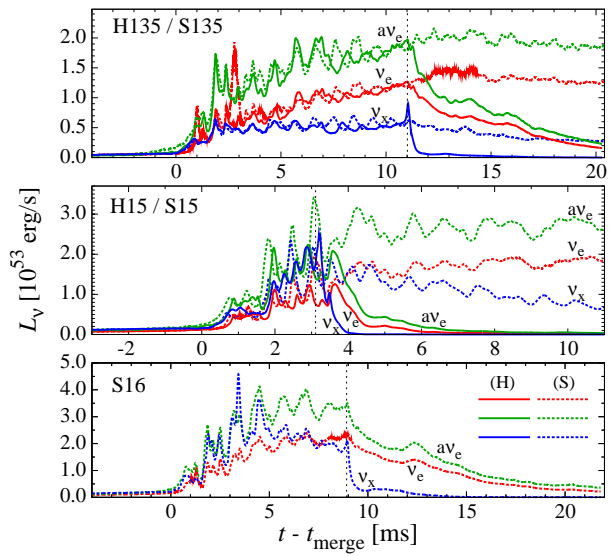}
 \vspace{-3mm}
 \caption{Neutrino luminosities for H135 and S135 (top),
   H15 and S15 (middle), and S16 (bottom). The dotted vertical lines show
   the time at which a BH is formed. The solid and dashed curves 
   correspond to the result with the Hyp-EOS and Shen-EOS, respectively.
   \label{fig3}}
\end{figure}

\begin{figure*}[t]
\epsfxsize=3.2in
\leavevmode
\epsffile{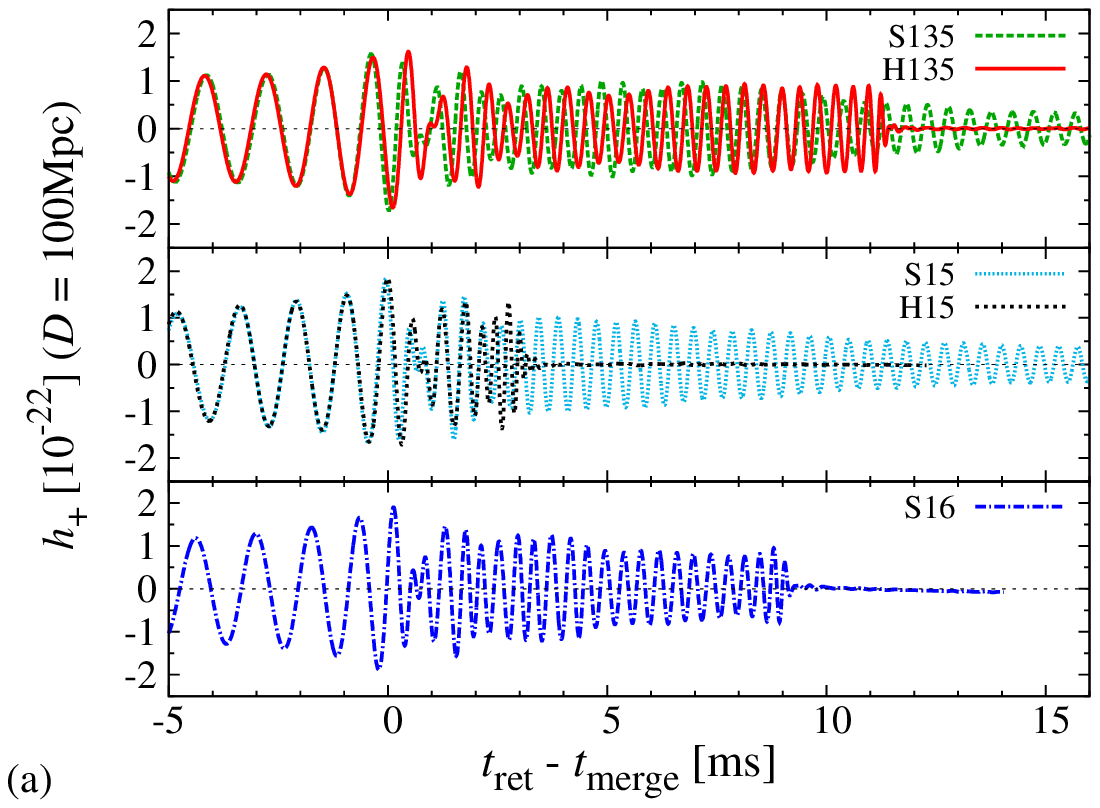}
\epsfxsize=3.2in
\leavevmode
~~~~~~~~\epsffile{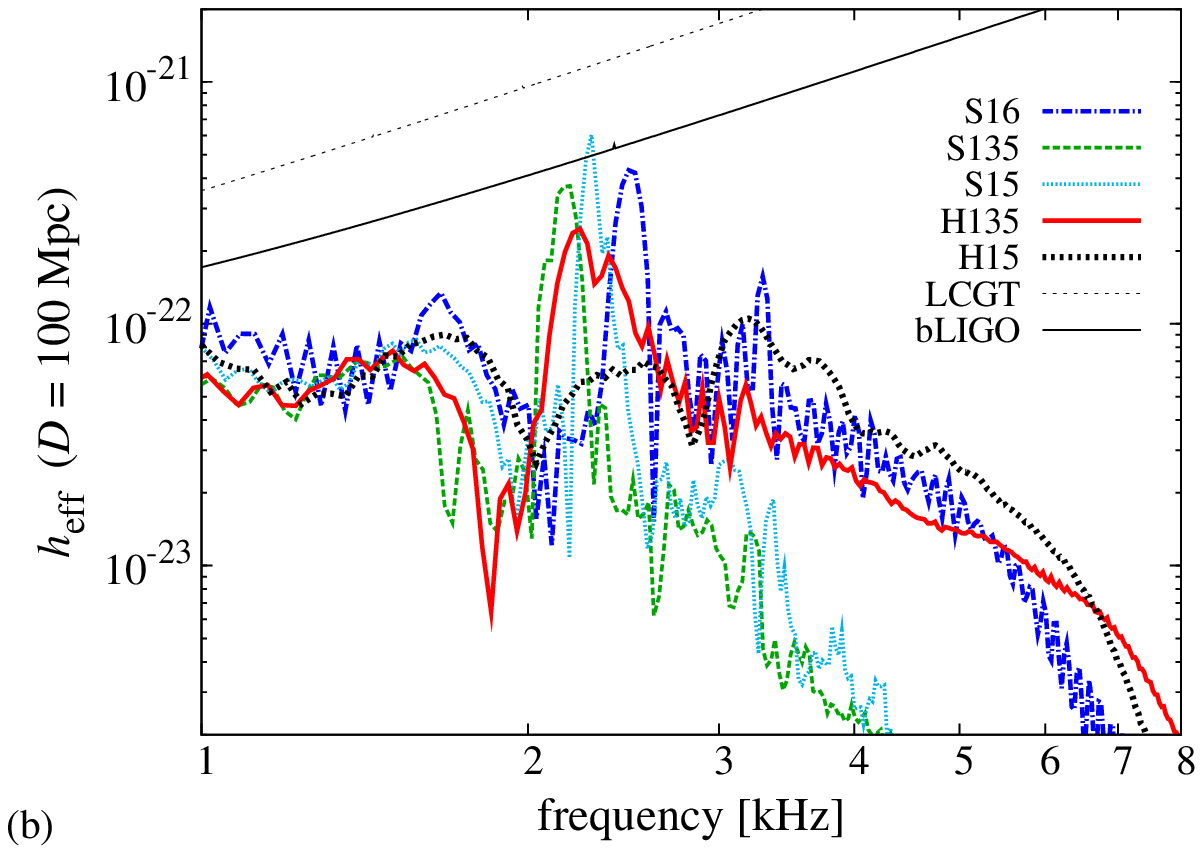}
\vspace{-2mm}
\caption{(a) GWs observed along the
axis perpendicular to the orbital plane for the hypothetical distance
to the source $D=100$~Mpc. (b) The effective amplitude of
GWs defined by $0.4f|h(f)|$ as a function of frequency for $D=100$~Mpc.
The noise amplitudes of a broadband configuration of
Advanced Laser Interferometer Gravitational wave Observatories (bLIGO),
and Large-scale Cryogenic Gravitational wave Telescope (LCGT)
are shown together.
\label{fig4}}
\end{figure*}

\begin{figure}[th]
 \epsfxsize=3.2in \leavevmode \epsffile{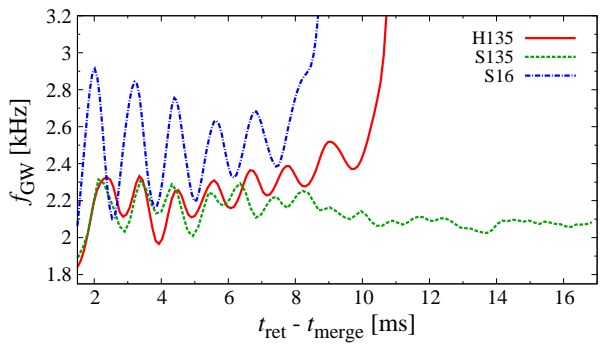} \vspace{-3mm}
 \caption{$f_{\rm GW}(t)$ in the HMNS phase, smoothed by a weighted
 spline, for H135 (solid red), S135 (dashed green), and S16 (dashed-dotted blue).
   \label{fig5}}
\end{figure}

{\em Numerical results}: Figure \ref{fig1} plots the evolution of
maximum rest-mass density, $\rho_{\rm max}$, maximum temperature,
$T_{\rm max}$, and maximum mass fraction of hyperons, $X_{\Lambda, {\rm max}}$
as functions of $t-t_{\rm merge}$ where $t_{\rm merge}$ is an
approximate onset time of the merger.  Before the merger ($t<t_{\rm
merge}$), $\rho_{\rm max}$ and $T_{\rm max}$ for H135 and H15 agree
well with those for the corresponding Shen-EOS model S135 and S15,
because $X_{\Lambda, {\rm max}}$ in this phase is small as
$O(10^{-2})$ and effects of hyperons on dynamics are not significant.
After the merger sets in ($t>t_{\rm merge}$), on the other hand, 
$X_{\Lambda,{\rm max}}$ increases to be $\gtrsim 0.1$ in accordance with the
increase in $\rho_{\rm max}$, and hyperons play a substantial role in
the post-merger dynamics and emission of GWs (see below).

Although the total mass, $M$, is larger than the maximum mass of
the zero-temperature spherical NS for all the models, a HMNS is formed
after the merger, supported by the centrifugal force and thermal
contributions to the pressure~\cite{Sekiguchi1,Hotokezaka}. They
subsequently contract by emission of GWs, and for H135, H15, and S16,
they collapse to a black hole (BH) at $t=t_{\rm BH}$ where
$t_{\rm BH}-t_{\rm merge} \approx 11.0$, $3.1$, and $8.9$ ms,
respectively.  The HMNS for S135 and S15 will not collapse to a BH in
a cooling time, $t_{\rm cool} \equiv E_{\rm th}/L_{\nu} \sim 2$--3 s,
where $E_{\rm th}$ is total thermal energy and $L_{\nu}$ is total
neutrino luminosity~\cite{Sekiguchi1}.  After the BH formation for
H135, H15, and S16, an accretion torus is formed around the BH. Figure
\ref{fig2} plots the torus mass, $M_{\rm torus}$, as a function of
time. $M_{\rm torus}$ in a quasistationary state is $\approx 0.082$,
0.035, and $0.10M_{\odot}$ for H135, H15, and S16, respectively, with
$\rho_{\rm max} \approx 10^{13}$ g/cm$^{3}$ (see Fig.~\ref{fig1}). For H15,
$M_{\rm torus}$ is smaller than those for other models because angular
momentum is not significantly transported in the outer region of the
HMNS due to its short lifetime.  Interestingly, $M_{\rm torus}$ is
smaller for H135 than that for S16 in spite of the longer lifetime of
the HMNS. This is likely because the HMNS for H135 just before the BH
formation is more compact than that for S16. Our results suggest that
the emergence of non-nucleonic degrees of freedom such as hyperons
would have a negative impact for driving a short-gamma ray burst for
which formation of a massive torus is preferred~\cite{GRB-BNS}.

Figure \ref{fig3} plots neutrino luminosities as functions of time.
Electron antineutrinos are dominantly emitted for all the models as found
in the previous studies~\cite{Ruffert,Sekiguchi1}.  For H135, H15, and
S16, the $\mu/\tau$ neutrino luminosity shows a spike around $t\approx
t_{\rm BH}$ because high temperature is achieved as a result of
compression.  Although $T_{\rm max}$ is very large as $\gtrsim 100$
MeV, the peak amplitude is moderate because of long diffusion
timescale of neutrinos due to the high density.  Soon after the BH
formation, $\mu/\tau$ neutrino luminosity steeply decreases because
high temperature regions are swallowed into the BH, while luminosities
of electron neutrinos and antineutrinos decrease only gradually
because these neutrinos are emitted via charged-current processes from
the massive accretion torus \cite{Sekiguchi1}.  These features in
neutrino luminosities are quantitatively the same for Hyp-EOS and
Shen-EOS models, and hence, it would be difficult to extract
information of the NS matter only from the neutrino signal.

Figure \ref{fig4}(a) plots the plus mode ($h_{+}$) of GWs as a
function of $t_{\rm ret}-t_{\rm merge}$ where $t_{\rm ret}$ is the
retarded time, extracted from the metric in the local wave zone.  The
GW amplitude is $h_{+} \lesssim 2\times 10^{-22}$ for a source at a
distance $D=100$ Mpc. GWs from the inspiral phase (for $t_{\rm ret}
\lesssim t_{\rm merge}$) agree well with each other for the models
with Hyp-EOS and Shen-EOS for the same mass. On the other hand,
quasi-periodic GWs from the HMNS (for $t_{\rm ret} \gtrsim t_{\rm
merge}$) show differences.  First, the amplitude of
quasi-periodic GWs damps steeply at the BH formation for H135 and
H15. This is because the HMNS collapse to a BH due to the softening of the EOS 
before relaxing to a stationary spheroid.  
Second, the characteristic GW frequency, $f_{\rm GW}$, {\em increases} with 
time for Hyp-EOS models.  These facts are
clearly observed in the effective amplitude [see Fig.~\ref{fig4}(b)]
defined by $h_{\rm eff}
\equiv 0.4 f |h(f)|$ where $h(f)$ is the Fourier transform of $h_+ -i
h_{\times}$ with $h_{\times}$ being the cross mode and the factor 0.4
comes from taking average in terms of direction to the source and
rotational axis of the HMNS.  Reflecting a shorter lifetime of the
HMNS in Hyp-EOS models, the peak amplitude of $h_{\rm eff}(f)$ is
smaller, in particular for H15 where the HMNS survives only for a
short period $\sim 3$~ms.  Reflecting the shift of the characteristic
frequency, the prominent peak in the GW spectra for Hyp-EOS
models (H135 and H15) is broadened. We here note that the lifetime 
of the HMNS is longer for H135 than that for S16.
The reason for this broadening of the peak is described as follows 
in more detail.

In the case that hyperons are absent, the HMNS slightly contract
simply due to the angular momentum loss (weakening centrifugal force) via GW emission.
By contrast, in the case that hyperons are present, $X_{\Lambda}$
increases with the contraction of the HMNS, resulting in the relative
reduction of the pressure. As a result, the HMNS further contract.
Recent studies showed that $f_{\rm GW}$ is associated with the
frequency of $f$-mode which is approximately proportional to
$\sqrt{M_{\rm H}/R_{\rm H}^{3}}$ where $M_{\rm H}$ and $R_{\rm H}$ are
the mass and radius of the HMNS~\cite{oscillation,BJ11}. This
indicates that $f_{\rm GW}$ should increase with time.  To see that
this is indeed the case, we show $f_{\rm GW} (\equiv d\phi_{\rm NP}/dt)$ 
calculated from a Weyl scalar $\Psi_{4} \equiv  \Psi e^{-i\phi_{\rm NP}}$ 
in the HMNS phase for H135, S135, and S16 in Fig.~\ref{fig5}.
It is clearly seen that the mean value of $f_{\rm GW}$ in the HMNS
phase is approximately constant for Shen-EOS models; $f_{\rm GW}
\approx 2.1$ and 2.5 kHz for S135 and S16, respectively.  By contrast,
$f_{\rm GW}$ for H135 increases with time (from $f_{\rm GW}\approx
2.0$ kHz at $t_{\rm ret}-t_{\rm merge}=2$ ms to $\approx 2.5$ kHz at
$t_{\rm ret}-t_{\rm merge}=10$ ms) as the HMNS becomes compact.

Our results raise a caution that one should be careful when using the
peak frequency of the GW spectrum to extract information of the NS
matter, because $f_{\rm GW}$ evolves with time for Hyp-EOS models.
This makes it ambiguous to determine $f_{\rm GW}$ and to relate
$f_{\rm GW}$ with the HMNS structure.  Our results rather suggest a
possibility that the emergence of hyperons may be captured from the 
evolution of the characteristic frequency of GW and the peak width of 
the GW spectra. Accompanied with this broadening, the peak amplitude
in $h_{\rm eff}$ decreases by a factor of a few.
%It should be also addressed that this
%{\bf could} be also applied for the case that {\bf {\it the other}}
%{\it non-nucleonic degrees of freedom} emerge and the corresponding 
%EOS becomes softer.
%Furthermore, observations of GW spectra around $f=f_{\rm GW}$ {\bf might}
%provide a potential opportunity to extract information of the phase
%transition between nucleonic and non-nucleonic phases, especially the
%deconfined quark phase: For example, suppose that mixed phases do not
%exist between the nucleonic and non-nucleonic phases. Then, $f_{\rm
%GW}$ will not change continuously but may show a jump at the phase
%transition, and two distinct peaks, associated with the nucleonic and
%quark phases, may appear in the GW spectra.

{\em Summary}: We have reported effects of hyperons on the BNS merger
in numerical-relativity simulations incorporating finite-temperature,
both hyperonic and nucleonic EOS, microphysical processes, and
neutrino cooling.  We showed that for the adopted hyperonic EOS, a BH
is not promptly formed and a HMNS is first formed for the typical
total mass of BNS ($\approx 2.7M_{\odot}$), as in the nucleonic
Shen-EOS~\cite{Sekiguchi1}.  The HMNS subsequently collapse to a BH
and a massive torus is formed around the BH.  The torus mass for the
hyperonic EOS is smaller than that for the nucleonic EOS.  We further
showed that for the hyperonic EOS, the characteristic frequency of
GWs, $f_{\rm GW}$, from the HMNS increases by a factor of $\sim
20$--30\% during their evolution via GW emission, by contrast with that for the
nucleonic EOS in which $f_{\rm GW}$ is approximately constant. 
Although the hyperonic EOS adopted in this {\it Letter} has some limitations that
it only takes $\Lambda$ hyperons into account and that it cannot produce a
stable NS with mass $M_{\rm J1614-2230}$, we believe from
our results that it will be possible to constrain the composition of
the NS matter via observation of GW from the HMNS.

{\em Acknowledgments}: Numerical simulations were performed on SR16000
at YITP of Kyoto University and on SX9 and XT4 at CfCA of NAOJ.  This
work was supported by Grant-in-Aid for Scientific Research (21018008,
21105511, 21340051, 22740178, 23740160), Grant-in-Aid on Innovative
Area (20105004), and HPCI Strategic Program of Japanese MEXT.

\end{document}